\documentclass{aastex63}
\usepackage{amsmath}
\usepackage{amssymb}
\usepackage{amsfonts}
\usepackage{wasysym}
\usepackage{labelschanged}

\usepackage{listings, color}    
\usepackage{textcomp}
\usepackage{fancyvrb}
\usepackage{verbatim}
\usepackage{sverb}

\RecustomVerbatimCommand{\VerbatimInput}{VerbatimInput}%
{fontsize=\footnotesize
}

\usepackage{tikz}
\usepackage{tikzscale}
\usepackage{tikz-3dplot}
\usepackage{pgfplots}
\usetikzlibrary{shapes}
\usetikzlibrary{shadings}
\usetikzlibrary{positioning}
\usetikzlibrary{calc,3d,shapes, pgfplots.external}
\usetikzlibrary{external}
\usepgfplotslibrary{external} 
\pgfplotsset{compat=newest}

\usepackage{hyperref}
\usepackage{ifthen}

\graphicspath{{./picts/}} %Setting the graphicspath

\newcommand{\bea}{\begin{eqnarray}}
\newcommand{\eea}{\end{eqnarray}}
\newcommand{\be}{\begin{equation}}
\newcommand{\ee}{\end{equation}}
\newcommand{\mb}{\mathbf}
\newcommand{\LongSheet}{\textit{``long sheet" }}
\newcommand{\ShortSheet}{\textit{``short sheet" }}
\newcommand{\LDK}{LDK2020}
\makeatletter
\def\input@path{{./picts/}}
\makeatother

\makeatletter
\newcommand{\gettikzxy}[3]{%
  \tikz@scan@one@point\pgfutil@firstofone#1\relax
  \edef#2{\the\pgf@x}%
  \edef#3{\the\pgf@y}%
}
\makeatother

\usepackage{math}
\usepackage{conventions}

\begin{document}

\title{The Onset of Magnetic Reconnection in Dynamically Evolving Coronal Current Sheets}
\author [0000-0003-0072-4634]{James E. Leake}
\affiliation{Heliophysics Science Division, NASA Goddard Space Flight Center 8800 Greenbelt Rd., Greenbelt, MD 20771, USA}
\author	[0000-0002-1198-5138]{Lars K.~S.~Daldorff}
\affiliation{Heliophysics Science Division, NASA Goddard Space Flight Center 8800 Greenbelt Rd., Greenbelt, MD 20771, USA}
\affiliation{Catholic University of America, Washington, DC, USA}
\author [0000-0003-2255-0305]{James a. Klimchuk}
\affiliation{Heliophysics Science Division, NASA Goddard Space Flight Center 8800 Greenbelt Rd., Greenbelt, MD 20771, USA}

\begin{abstract}

We present the first results of three-dimensional (3D) numerical magnetohydrodynamic (MHD) simulations of the onset of magnetic reconnection via the tearing instability in dynamically thinning current sheets in the solar corona. In all our simulations, the onset of the non-linear tearing instability, which leads to the break-up of the thinning current sheet, does not occur until after the instability growth time becomes faster than the dynamic thinning time. Furthermore, as in previous 3D MHD simulations of static current sheets in the corona, for some parameters, the amount of magnetic shear is a fundamental switch-on parameter, which has consequences for coronal heating models. These results open up the possibility of using observable quantities of coronal current sheets to predict when they will break-up and release magnetic energy to power various energetic phenomena and/or heat the atmosphere.

\end{abstract}

\section{Introduction}

Magnetic reconnection is considered a fundamental process in astrophysics, and is often invoked as the process which releases energy, to both power the heating of the solar atmosphere \citep{Klimchuk_2015}, and to drive energetic events in the solar atmosphere, such as flares, jets, prominence eruptions, and coronal mass ejections \citep[e.g.][]{Karpen2012,Wyper2017}. This energy is released as kinetic and thermal energy, and energy which accelerates particles, but is stored as magnetic energy in the non-potential ($\J\ne0$) magnetic field that develops in the corona as a result of the combination of emerging magnetic fields and photospheric surface plasma flows. Magnetic reconnection in a conducting plasma occurs when magnetic field topology changes, and requires the build up of large gradients in this magnetic field, known as \textit{current sheets}. {\color{black} Understanding the onset of magnetic reconnection in the corona is of vital importance, as the timing of when the energetic release occurs affects the amount of free magnetic energy that can build up \citep{Klimchuk_2015}.}

{\color{black}
One complication with studying magnetic reconnection in the solar context is that the original standard Sweet-Parker model \citep{Sweet1958,Parker1957} is too slow to account for the fast reconnection believed to be happening in solar flares, as it predicts that the reconnection rate goes as ${S_L}^{-\frac{1}{2}}$ where $S_L$ is the Lundquist number using the current sheet length $L$. However, \citet{Loureiro2007, Huang2013} showed that very thin current sheets are unstable and can result in localized shorter, thinner current sheets that result in fast reconnection, with the maximal growth rate scaling as ${S_L}^{1/4}$. This was termed the \textit{plasmoid instability} as the 2D evolution of this instability resembles plasmoid structures. Interestingly it was shown in \citet{Pucci2014, Pucci2017} that this plasmoid instability is merely the application of the scaling of the tearing instability \citep{Furth1963, Coppi1976} to sheets that are already scaled based on the S-P model ($a/L \sim S_{L}^{-\frac{1}{2}}$).

 For a complete study of the reconnection process, the detailed flow of energy from magnetic to a combination of kinetic, thermal, and particle acceleration necessitates a kinetic description of the plasma. in order to include both the large scale driving scales, and the detailed description of energy transfer at small scales. However, doing so is computationally prohibitive if one wants to simulate coronal scales, which are typically Mm, and still resolve the kinetic scales, which in the rarefied corona can be as low as cm. However, \citet{Pucci2014, Pucci2017} argued that a dynamically thinning system will undergo fast tearing (with the growth time comparable to an Alfv\'{e}n crossing time of the sheet) before it reaches an aspect ratio consistent with S-P reconnection, and that for the solar corona, this is approximately 300 m, which is above the kinetic scale. Hence, studying just the onset of the tearing instability and the break up of a current sheet with a fluid description, while missing some of the details of the reconnection once it is occuring, is a reasonable approach to investigating the onset of magnetic reconnection in the corona. 
}

As is well known, when observed in 2D, the tearing instability forms a chain of magnetic islands, which are the cross sections of flux tubes in 3D. Each tearing mode consists of identical tubes that are lined up side by side within a plane. The plane is parallel to the current sheet and can be located at the center of the sheet or offset from it. The tubes are aligned with the local magnetic field direction, and for a current sheet where the field rotates across the sheet, the tubes have a variable orientation. Modes at the center of the sheet ($y=0$, where $y$ is the direction across the current sheet) are known as parallel modes because the tubes are parallel to the guide field ($B_z$), while modes offset from the center are known as oblique modes. Each mode is characterized by wave-numbers $k_x$ and $k_z$, which are the inverse wavelength along the sheet and along the guide field, respectively. Sometimes $m$ and $n$ are used, being the number of modes in each direction.

Some theoretical studies of the tearing instability in developing and/or thinning current sheets have been performed. \citet{Tollman2013} identified three phases of development for a current sheet formed by shearing flows: The first being the stage where the tearing growth time is much larger than the dynamical driving timescale, and so little growth occurs relative to the external forcing. The second phase is when the growth timescale and the dynamical timescale are comparable, and so separating the temporal dependence of the linear tearing instability was not possible. The last stage is when the growth timescale is much faster than the dynamical timescale, and the system is effectively like a static sheet.

\citet{Uzdensky2016} showed that during the collapse of a current sheet's width, the non-linear phase is dictated by the mode or modes that are dominant during the linear phase, and as the number of modes with significant growth rate increases with decreasing width, the number of X points in the non-linear stage along a current sheet depends not only on the sheet parameters but appears to depend on the thinning rate. However, a detailed investigation of this process is necessary to determine the nature of the break-up of a thinning current sheet. To date, no analysis of the tearing instability has been performed on 3D MHD simulations of dynamically evolving current sheets. {\color{black} As discussed in \citet{Tenerani2015} there are instances of evolving current sheets in simulations of the solar corona and elsewhere \citep[e.g.,][]{Rappazzo2013,Grauer2000,Bratchett2013}, but they are not resolved sufficiently for detailed investigation, and so the transition from linear to non-linear tearing is not properly simulated. To achieve this, simplified geometries are required to be studied first.}

{\color{black} \citet{Comisso2017} derived time dependent scaling laws for magnetic reconnection, relating the growth rate to the Lundquist number, and found that the discrepancy of those laws from the standard tearing scaling laws depended on Lundquist and magnetic Prandtl (ratio of viscous to resistive forces), the nature of the noise used to seed the instability, and the thinning process, the last of which can be very important.}

{\color{black} Previous work by \citet{Tenerani2015} reported results of 2D MHD simulations of thinning current sheets, where the background current sheet was modified in time with a exponential decrease of timescale $\tau_c$ The fastest growing parallel mode (no oblique modes were present in 2D) and the observed growth rate matched the time dependent theory in the linear stages of the tearing instability, and they found that onset of the tearing instability occurred when the fastest (over all possible modes) growth rate $\tau_m=\tau_c$. They also reported that fast reconnection occurred before the current sheet reached a Sweet-Parker scaling, as predicted by \citet{Pucci2014, Pucci2017}.}

In a recent paper \citep[called \LDK ~ hereafter,][]{Leake2020}, the onset of magnetic reconnection was investigated in static 3D coronal current sheets, with a small parameter study consisting of various values of the magnetic shear and current sheet length. Two regimes in this parameter study were identified. The first, called the \LongSheet regime, occurred when the dominant wavelength $\lambda_x$ of the 3D tearing instability in the direction along the sheet was smaller than the length of the current sheet, i.e. $\lambda_x = 2\pi/k_x < L_x$ or $m = L_x/\lambda_x > 1$. In this regime,
the onset of magnetic reconnection was dominated by the growth and coalescence of the sub-harmonics of this dominant mode. The outflows of these coalescing modes drove shock heating. Hence, in this regime, the onset and rate of reconnection were only weakly dependent on the magnetic shear ($B_x$) of the sheet, because the value of the shear mainly affects the strength of the oblique modes, and the coalescence of the parallel mode evolution dominates the reconnection and heating.

The second regime in \LDK ~ was called the \ShortSheet regime, where $\lambda_x = 2\pi/k_x > L_x$ or $m = L_x/\lambda_x < 1$. Thus the linear evolution was dominated by a single parallel mode, which was unable to coalesce with itself, and had no sub-harmonics in the corresponding infinite system. In this regime, the strength of the shear field was very important: {\color{black} For strong shear, there was an oblique mode (1,1) of the same wavelength in the x-direction and a similar growth rate as the fastest growing  parallel mode (1,0)}, which was able to interact with the parallel mode and drive turbulent-like reconnection. For weak shear, the oblique modes were much weaker than the parallel modes, and the dominant parallel mode was unable to drive significant reconnection, and ultimately saturated. Thus, in this regime, the magnetic shear was identified as a possible switch-on for fast reconnection in the solar corona. In this paper, it will be argued that the \LongSheet regime applies to the magnetic reconnection associated with eruptive flares in the solar corona, and the \ShortSheet regime applies to reconnection associated with nanoflares. Nanoflares are smaller in energy, but believed to heat the corona due their increased frequency. The importance of magnetic shear in the heating of the corona was raised initially by \citet{Parker1988}, and is discussed in \citet{Klimchuk_2015}

The simulations in this paper are designed to answer two main questions. Firstly, what determines when a thinning current sheet becomes unstable to the 3D non-linear tearing instability on a dynamically important timescale? Secondly, does the above result from static simulations hold when the system is dynamically thinning (i.e. is shear a possible switch-on mechanism, depending on the regime)? In \S\ref{sec:theory}, we perform multiple thought experiments of thinning current sheets in the corona, designed to demonstrate that indeed, eruptive flares are in the \LongSheet regime, and nanoflares are in the \ShortSheet regime. We then use these thought experiments to design 3D MHD simulations that investigate the importance of magnetic shear in these two regimes. \S\ref{sec:numerics} describes the numerical MHD code used for these simulations. In \S\ref{sec:3D} we perform and analyze those simulations, focusing on the point at which the thinning system breaks up via the non-linear tearing instability, and the nature of that breakup in the context of the above mentioned \LongSheet and \ShortSheet regimes. In \$\ref{sec:2D} we perform a small 2D parameter study to show that the behavior in the 3D study is robust when considering different dynamical thinning timescales and coronal diffusivities. Finally, we conclude with contextual discussion and a description of the caveats of this work in \S\ref{sec:discussion}.

\section{Theoretical study of  thinning coronal current sheets}
\label{sec:theory}

Before looking at the 3D MHD simulations, it is useful to perform some thought experiments of some idealized thinning current sheets, in order to relate the MHD simulations performed here to both eruptive flares and nanoflares, and also to  place the simulations in 
context with the static simulations of \LDK. 

Let us consider a  current sheet with initial width $a$, length $L_x$, in a coronal plasma of resistivity $\eta$, diffusivity $D=\eta/\mu_0$, and local Alfv\'{e}n speed $V_a$. The maximal growth rate of the tearing instability scales as \citep{Baalrud2012,Pucci2014}
\begin{equation}
    \gamma_{g}\tau_a \sim {S_a}^{-1/2},
    \label{eqn:tau_g_orig}
\end{equation}
where $\tau_a=a/V_A$ and $S_a=aV_A/D$. This can be rearranged in terms of the tearing growth timescale $(\tau_g = 1/\gamma_g)$:
\begin{equation}
    \tau_g \sim {a}^{\frac{3}{2}} {V_{a}}^{-\frac{1}{2}} {D}^{-\frac{1}{2}}. \label{eqn:tau_g}
\end{equation}  
Similarly, the wavenumber ($k$) of the fastest growing mode of the tearing instability scales as
\begin{equation}
    k a \sim {S_{a}}^{-\frac{1}{4}},
\end{equation}
which can be rearranged in terms of a wavelength ($\lambda = 2\pi/k)$:
\begin{equation}
    \lambda \sim 2\pi {a}^{\frac{5}{4}} {V_{a}}^{\frac{1}{4}} {D}^{-\frac{1}{4}} \label{eqn:lambda}.
\end{equation}
Hence, with all other variables the same, if the width of the current sheet $a$ decreases with time, the growth timescale ($\tau_g$) decreases (the instability gets faster), and the dominant mode has a smaller and smaller wavelength. Furthermore, imagine a current system with a large enough width that the instability growth timescale is much slower than the dynamical timescale ($\tau_t$) which thins the sheet in time. At some point, the growth timescale will become faster than the thinning timescale. As discussed in \citet{Tollman2013}, when the growth timescale is comparable to the thinning timescale, the instability is tearing the sheet as it is being thinned, and when the growth timescale is
much faster than the thinning timescale, the instability effectively ``sees" a static background. At some point during this transition from $\tau_g>\tau_t$ to $\tau_g<\tau_t)$ we expect the system will become unstable to the non-linear tearing instability. Let us assume that this happens close to the ``cross-over'' point $\tau_g = \tau_t$. When this happens, one can  estimate what the width of this sheet will be, by setting $\tau_g = \tau_t$ in Eqn. (\ref{eqn:tau_g}) and rearranging:
\begin{equation}
    a(\tau_g\sim\tau_t) \sim {\tau_t}^{\frac{2}{3}}{V_{a}}^{\frac{1}{3}}{D}^{\frac{1}{3}}.
\end{equation}
One can then plug this into  Eqn. (\ref{eqn:lambda}) to get
\begin{equation}
    \lambda(\tau_g\sim\tau_t) \sim 2\pi{\tau_t}^{\frac{5}{6}}{V_A}^{\frac{2}{3}}{D}^{\frac{1}{6}}.
\end{equation}
Hence, when a system thins to the point at which $\tau_g=\tau_t$ one can place this time-dependent situation in terms of the two \LongSheet and \ShortSheet regimes in the static simulations of \LDK. Specifically, the system is in the \LongSheet regime if $\lambda(\tau_g\sim\tau_t) < L_x$ or $m = L/\lambda(\tau_g\sim\tau_t)>1$, and is in the \ShortSheet regime if $\lambda(\tau_g\sim\tau_t) > L_x$, or $m = L/\lambda(\tau_g\sim\tau_t)<1$. 

Having obtained this relationship between $\lambda(\tau_g\sim\tau_t)$ and the current sheet parameters of thinning timescale, Alfv\'{e}n speed, and the plasma diffusivity, we can now first choose some values relevant to the solar corona, particularly eruptive flares and nanoflares, and then choose values that can be accurately simulated using MHD simulations.

However, the simple scaling laws used above, Eqns. (\ref{eqn:tau_g},\ref{eqn:lambda}) are asymptotic values assuming a continuous spectrum of available tearing modes, and do not take into account the available parallel and oblique modes of a 3D current sheet system such as the one used in the MHD simulations here. These are restricted by the periodic boundary conditions in the simulations and by finite current sheet size in real systems.

{\color{black} For a Harris current sheet $B_{x}(y) = B_{x,0}\tanh{(y/a)}$ with a guide field $B_z=B_{z,0}$, and for available wave-numbers $k_x,k_z$ in the $x$ and $z$ directions,  the linear tearing instability \citep{Baalrud2012} has growth rate}
\begin{eqnarray}
    \gamma \tau_a \sim 
    \begin{cases}
        {S_{a}}^{-\frac{3}{5}}{(ka)}^{-\frac{2}{5}}{(1-\mu^2)}^{\frac{2}{5}}{(1+\mu^2-k^2a^2)}^{\frac{4}{5}}, ~ ka>{S_{a}}^{-\frac{1}{4}}{(1-\mu^2)}^{-\frac{1}{4}}{(1+\mu^2)}^{\frac{3}{4}} \\
        {S_{a}}^{-\frac{1}{3}}{(ka)}^{\frac{2}{3}}{(1-\mu^2)}^{\frac{2}{3}} ~ ~ ~ ~ ~  ~ ~ ~ ~ ~ ~ ~ ~ ~ ~ ~ ~ ~  ~ ~ ~ ~ ~ ~ ~ ~ ka<{S_{a}}^{-\frac{1}{4}}{(1-\mu^2)}^{-\frac{1}{4}}{(1+\mu^2)}^{\frac{3}{4}}
        \end{cases}
        \label{eqn:full_growth}
\end{eqnarray}
{\color{black} where $k=|(k_x,k_z)|$ and $\mu=k_zB_{z,0}/k_xB_{x,0}$. The two cases are for the constant $\psi$ and non-constant $\psi$ regimes \citep{Coppi1976}. The intersection of the two represents the maximum growth rate where $\gamma_g\tau_a \sim S_{a}^{-1/2}$ as in the simple expression (\ref{eqn:tau_g_orig}). Note that going forward, we will calculate the theoretical growth rate for a thought experiment
or simulation as follows: We take, at each time, the current sheet width, strength ($B_{x,0}$ and $B_{z,0}$), and average gas density, either from the thought experiment parameters or using a fit of the original current sheet profile in the current simulation data. We then calculate,  for most available modes ($k_x,k_z$) or ($m,n$) in the system of size $L_x$, the growth rate using Eqn. (\ref{eqn:full_growth}). We typically choose modenumbers $(m,n)$ in the ranges $m \in [0,50]$ and $n \in [-25,25]$, but only consider the subset of these choices which have significant growth rates). Then, for determining whether the system is in the \ShortSheet or 
\LongSheet regimes, we calculate $m=\max{(1,L_x/\lambda_{max})}$ where $\lambda_{max}$ is the wavelength in the $x$ direction of the fastest available mode. Note that at various points in this paper we will calculate the discrete modenumber $m$ of the fastest growing mode at a particular time, such as when $\tau_g=\tau_t$, which we denote as e.g., $m(\tau_g=\tau_t)$.}

%or simulation using (\ref{eqn:full_growth}) for the available wavenumbers in the system being considered, with estimates of the current sheet strength ($B_{x,0}$ and $B_{z,0}$)
%and width. }

%{\color{black}
%Note that for the 2D case where there are no oblique modes ($k_z=0$), then this simplifies to $\gamma_g\tau_a \sim {S_a}^{-\frac{1}{2}}$ and $k_{max}a\sim{S_a}^{-\frac{1}{4}}$ which is the simple calculation used above. }

%Now we can calculate the tearing growth timescale for a dynamically evolving sheet, calculate the wavelength of the dominant 3D mode that fits discretely into the finite length $L_x$. We can then calculate the discrete mode-number of this dominant mode as $m=\max{(1,L/\lambda)}$. We can then work out which regime these sheets lie in at the point of cross-over $m(\tau_g=\tau_t)$ and at a later time $m(\tau_g=\tau_t/50)$ when the tearing timescale is much faster than the thinning timescale.

\begin{table}[h]
\begin{center}
\begin{tabular}{ |c|c|c|c|c|c|c||c|c|}
\hline
 Name & $\tau_{t} (s)$ & $a_0 (km)$ & $D (m^2/s)$ & $L_x (Mm)$  & $S_{a}$ & $S_{L}$ & $m(\tau_g=\tau_{t})$ & $m(\tau_g=\frac{\tau_{t}}{50})$ \\ 
 \hline
 Nanoflare & 1000 & 10 & 1 & 0.1 & $2.3\times10^{10}$& $2.3\times10^{11}$& 1 & 1  \\
  \hline
 Eruptive flare & 100 & 50 & 1 & 10 & $1.15\times10^{11}$& $2.3\times10^{13}$& 2 & 52 \\
 \hline
  \hline
Short sheet sim & 1000 & 1e4 & 1e8 & 20 &  $2.3\times10^{5}$& $4.6\times10^{5}$&  1 & 1   \\
  \hline
Long sheet sim & 1000 & 1e4 & 1e8 & 100 &  $2.3\times10^{5}$& $2.3\times10^{6}$&  1 & 4   \\
\hline
\end{tabular}
\end{center}
\caption{Relevant parameters of 4 thought experiments }
\label{table:calc}
\end{table}

Consider the situation of \textit{nanoflares}, which are believed to be driven by photospheric motions entangling magnetic fields in the corona creating dynamically thinning current sheets. The thinning timescale thus is taken to be  a convective turnover time 1000 s. We use a realistic diffusivity of the corona of $1 ~ \textrm{m}^{2}/\textrm{s}$, an initial width of 10 km, a sheet length of $L_x=100$ km, corresponding to the characteristic diameter of an elemental coronal flux tube \citep{Klimchuk_2015}, and a local Alfven speed of $1.6\times10^{6} ~ \textrm{m}/\textrm{s}$. This leads to $m(\tau_g\sim\tau_t)=m(\tau_g=\tau_t/50) = 1$ and so this thought experiment for a nanoflare is in the \ShortSheet regime for all of its evolution.

Now consider the situation of an eruptive flare.
Flare current sheets are produced during the initial stages of a coronal mass ejection.  Assuming that the sheet thins at the same rate that it lengthens during the eruption, and taking an initial length and eruption velocity of 10,000 km and 100 km/s \citep{Zhang2001}, we estimate a thinning time scale of $\tau_t=100 ~ \textrm{s}$.
Taking a longer length associated with the current sheet of an eruptive flare $L_x=100$ Mm, this leads to $m(\tau_g\sim\tau_t)=2$ and $m(\tau_g\sim\tau_t/50)=52$, so the system is marginally in the \LongSheet regime when $\tau_g\sim\tau_t$ and then quickly becomes more entrenched in this regime as the system thins. 

\begin{figure}[h]
    \centering
    \includegraphics[width=0.45\textwidth]{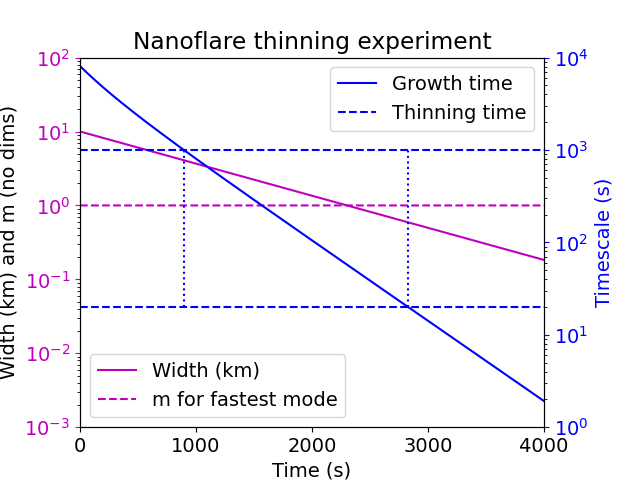} 
    \includegraphics[width=0.45\textwidth]{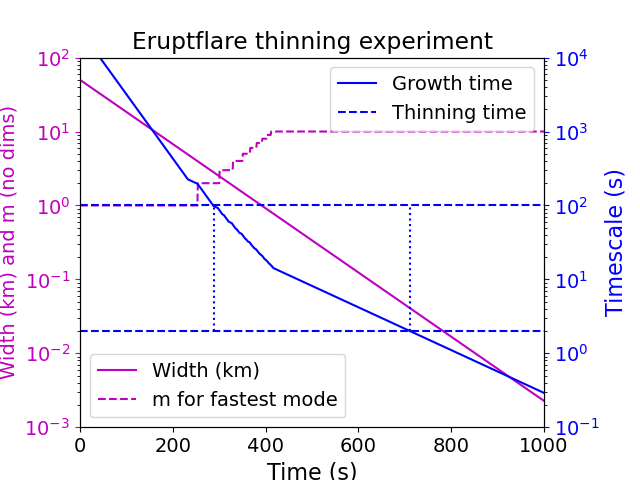} \\
    \includegraphics[width=0.45\textwidth]{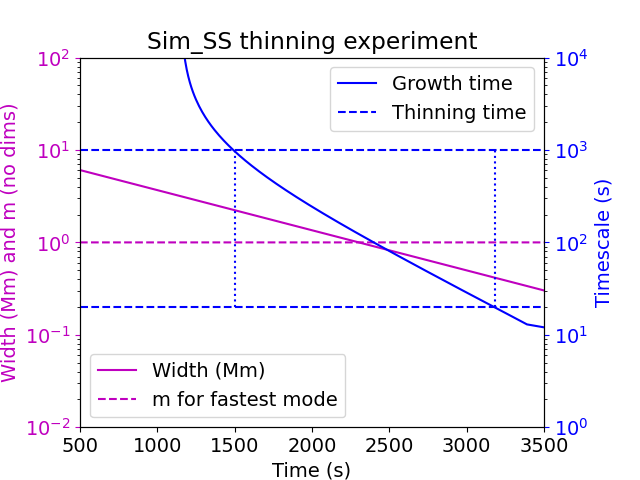} 
    \includegraphics[width=0.45\textwidth]{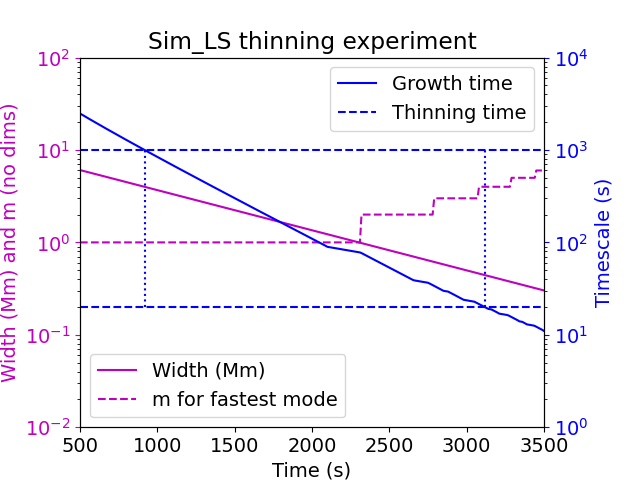} \\
    \caption{Results of idealized thinning currents. The top two panels show two different thinning current sheet scenarios, one representing a nanoflare current sheet (left), and one representing an eruptive flare (right). The bottom row shows a thinning current sheet based on the initial conditions of the 3D simulations below, and an idealized thinning rate. In all plots, the solid magenta line is the width of the sheet. The solid blue line is the timescale of the fastest growing available linear mode. The magenta dashed line is the mode-number in the $x$ direction (along the sheet). The two blue dashed lines represent $\tau_t$ and $\tau_t/50$, and the two vertical dotted lines show when $\tau_g$ equals $\tau_t$ and $\tau_t/50$. 
    \label{fig:theory_thinning}}
\end{figure}

{\color{black} The results of these two thought experiments are reviewed in Table \ref{table:calc} and the evolution in time is shown in the top two panels of Figure \ref{fig:theory_thinning}. The width of the sheet in the experiment is shown as a solid magenta line. At each time, the width, along with the sheet strength ($B_{x,0},B_{z.0})$ is used to estimate the growth rate of most available modes, using Eqn. (\ref{eqn:full_growth}), and then the fastest growing mode is found. The inverse of this maximum growth rate is plotted as the ``growth time", $\tau_g$, solid blue line. The horizontal dashed blue lines are at the thinning time $\tau_t$ and the value $\tau_t/50$. The vertical dotted lines are at times when $\tau_g=\tau_t$ and $\tau_g=\tau_t/50.$}

%A visual representation of the temporal evolution of both this eruptive flare thought experiment and the nanoflare one is shown in the top row panels of Figure \ref{fig:theory_thinning}, which show the relevant timescales, width, and the value of $m$ as a function of time.

Now we move onto designing MHD simulations that aim to represent similar evolution as the nanoflare and eruptive flare situations above. The coronal diffusivity of $1 ~ \textrm{m}^{2}/\textrm{s}$ cannot be achieved in simulations that also span many Mm, as the numerical diffusivity associated with practical resolution for these scales is much larger than the target value. Instead, we can use a practically resolvable  diffusivity of $D=10^{8} ~ \textrm{m}^{2}/\textrm{s}$, which is many orders larger than the corona. {\color{black} This results in vastly different Lundquist numbers for these proposed simulations compared to the thought experiments above (whether using the Lundquist number based on width, $S_a$ or based on length, $S_L$).}However, we can use different current sheet widths and lengths, and thinning timescales,  so that they fall into similar regimes as the above thought experiments. 
The initial width is $10$ Mm and the thinning timescale is 1000 s. As in \LDK, we use two values of the sheet length $L_x=20,100$ Mm. This leads to $m(\tau_g=\tau_{t})=1$ and $m(\tau_g=\frac{\tau_{t}}{50})=1$ for the choice of $L_x=20$ Mm, and so this simulation would lie in the \ShortSheet regime and is analogous to the nanoflare thought experiment above. This also leads to $m(\tau_g=\tau_{t})=1$ and $m(\tau_g=\frac{\tau_{t}}{50})=4$ for the choice of $L_x=100$ Mm, and so this simulation transitions from the \ShortSheet regime to the \LongSheet regime during the thinning process and  is analogous to the eruptive flare thought experiment above. {\color{black} The results for these proposed simulations are reviewed in Table \ref{table:calc}.}

Figure \ref{fig:theory_thinning} (bottom panels) shows the time-dependence of these two proposed MHD simulations. Note, that these plots assume the system evolves linearly for all times, and ignores any non-linear evolution. Based on this section, the 3D dynamically thinning MHD simulations in this paper are expected to lie in the long sheet regime for $L_x=100 ~ \textrm{m}$, and the short-sheet regime for $L_x= 20 ~ \textrm{Mm}$, as in the static cases for \citet{Leake2020}, and so, while having very different resistivities, are somewhat applicable to the case of eruptive flares (long-sheet) and nanoflares (short-sheet), respectively. It remains to be seen whether the 3D MHD simulations, driven to create thinning current sheets, will behave as expected, as parameters like local Alfv\'{e}n speed will change based on how the system is driven, and non-linear effects will come into play. A key question is also whether the onset of rapid tearing occurs after the cross-over, and by how much after, as this is potentially a prediction criteria that can be used when looking at the real corona, and when simulating current sheets in larger simulations, where the details of the onset may not be resolvable.

\section{Numerical Setup}
\label{sec:numerics}
\subsection{Equations and numerical parameters}

%subsection{Equations}
To study magnetic reconnection in the plasma of the solar corona in the fluid/MHD regime, the following visco-resistive MHD equations, written here in Lagrangian form are solved:

\begin{eqnarray}
\frac{D\rho}{Dt} & = & -\rho\nabla.\mathbf{V}, \\
\frac{D\mathbf{V}}{Dt} & = & \frac{1}{\rho}\left[-\nabla P
+ \mathbf{J}\times\mathbf{B} + \mb{F}_{shock}\right],\\
\frac{D\mathbf{B}}{Dt} & = & (\mathbf{B}.\nabla)\mathbf{V}
- \mathbf{B}(\nabla .\mathbf{V}) - \nabla \times (\eta\mathbf{j}), \\
\frac{D\epsilon}{Dt} & = & \frac{1}{\rho}\left[-P\nabla .\mathbf{V}
 + H_{visc} + \eta {J}^{2}\right],
\label{eqn:energy_MHD}
\end{eqnarray}
where $\rho$ is the mass density, $\mathbf{V}$ the velocity, $\mathbf{B}$ the magnetic field, and $\epsilon$ the internal specific energy density. The current density is given by $\mathbf{J}=\nabla\times\mathbf{B}/\mu_{0}$, where $\mu_{0}$ is the permeability of free space, and $\eta$ is the resistivity. The gas pressure, $P$, and the specific internal energy density, $\epsilon$, can be written as
\begin{eqnarray}
P & = & \rho k_{B}T/\mu_{m}, \\
\epsilon & =  & \frac{k_{B}T}{\mu_{m}(\gamma-1)}
\label{eqn:eos}
\end{eqnarray}
 respectively, where $k_{B}$ is Boltzmann's constant, $T$ is the temperature, and $\gamma$ is 5/3. The reduced mass $\mu_{m}$, for this fully ionized Hydrogen plasma, is given by $\mu_{m}=m_{p}/2$ where $m_{p}$ is the mass of a proton.

The equations are solved using a Lagrangian-Remap (LaRe) approach \citep{Arber2001}. 
The shock viscosity $\mathbf{F}_{shock}$ in the momentum equation is finite at discontinuities but zero for smooth flows, and the shock jump conditions are satisfied with an appropriate choice of shock viscosity \citep{Camanara1998,Arber2001}. This allows heating due to shocks to be captured as a viscous heating $H_{visc}$, while the Ohmic heating term is excluded from energy equation as the chosen resistivity is much larger than the real Sun \citep{Leake2020}.

The equations above are non-dimensionalized by dividing each variable ($C$) by its normalizing value ($C_{0}$). The set of equations requires a choice of three normalizing values. We choose normalizing values for the length, $L_{0}=10^{6}~ \textrm{m}$,  density, $\rho=1.5\times10^{-13} ~ \textrm{kg}/\textrm{m}^{3}$, and magnetic field, $B_{0}=10^{-3} ~ \textrm{T}$ ($10 ~ \textrm{G})$. This leads to a normalizing velocity of $V_{0} = 2.3\times10^{6} \textrm{m}/\textrm{s}$. All results presented in this study will use dimensional units.

\subsection{Initial Conditions and Dynamic Thinning}

To allow for rigorous validation and analysis of the numerical simulations, we use a simple initial equilibrium. The background density and temperature are $1.67\times10^{-13} ~ \textrm{kg}/\textrm{m}^{3}$ and $10^6 ~ \textrm{K}$. The magnetic field is a modified 1-D, force-free Harris current sheet with a guide field,
\begin{eqnarray}
B_{x}(y) & = & {B_{x,0}}\tanh{(\frac{y}{a})}\cos{(\pi \frac{y}{L_{y}})} \\
B_{z}(y) & = & \sqrt{(B_{z,0}^{2}-B_{x}(y)^{2})}
\end{eqnarray}
The cosine dependence on y reduces the shear component of the magnetic field ($B_{x}$) to zero at the ends ($\pm L_{y}/2$) and allows for the use of periodic boundary conditions in $y$ as well as the other two dimensions. $B_{x,0}$ and $B_{z,0}$ are the values far from the center of the sheet. They define a shear angle $\theta = 2 arctan(B_{x,0}/B_{z,0})$, which is the amount that the magnetic field rotates across the sheet. We choose a value of $B_{z,0}=10.06$ G for the guide field, and two values of the shear field $B_{x,0}=1.9,7.0$ G, which we call the \textit{weak shear} and \textit{strong shear} cases later in this paper. These choices result in corresponding angle of $\theta=21.4^{\circ},~70^{\circ}$. The initial width, as in the theoretical study above is 10 Mm, before this system is dynamically driven to thin the current sheet (see below).

To initiate the linear stage of the 3D tearing instability, the velocity $V_{y}$ is set to be a superposition of weak Fourier modes, with mode numbers $(m,l,n)$,
%$k_{x},k_{y},k_{z}$),
which encompass the predicted allowed dominant parallel and oblique modes in this particular system, based on the linear theory above\citep{Baalrud2012}:
 \begin{equation}
 V_{y}(x,y,z) = \sum_{m,l,n}{V_{0}}^{mln}{\sin{(k_{x}x-{\phi_{x}}^{mln})}\sin{(k_{y}y-{\phi_{y}}^{mln})}\sin{(k_{z}z-{\phi_{z}}^{mln})}}
 \end{equation}
where $k_{x} = 2\pi/\lambda_{x}=2\pi m/L_{x} ~( m=[1,n_m])$, $k_{y} = 2\pi/\lambda_{y} = 2\pi l /L_{y}~( l=[1,n_l])$, and $k_{z}=2\pi\lambda_{z} = 2\pi n/L_{z} ~(n=[0,n_n])$. To initiate a large range of modes we use $(n_m,n_l,n_n)=(25,5,5)$. The phases ($\phi^{mln}$) and amplitudes ($V_{0}^{mln}$) for each mode are chosen randomly, with the amplitudes chosen randomly from a top-hat function in the range [-10, 10] m~s$^{-1}$ This results in an average initial velocity of 43 m~s$^{-1}$.

%Despite the use of this specific initial condition on $V_{y}$ we have performed tests with different values of $n_{n,m,l}$ and also using randomly seeded values of $V_{y}$, and find that the evolution of the system is very similar, as the fastest growing modes allowed by the system quickly dominate the dynamics mostly  independent of the initial $V_{y}$ distribution. 

The simulation grid extends [-25,25] Mm in the $z$ direction (along the guide field), and [-400,400] Mm in the $y$ direction, across the current sheet. The large value of $L_y$ is to accommodate the forcing that we employ to drive the thinning of the current sheet (see below). Based on the theoretical study above, {\color{black} we choose two values of $L_x=20,100$ Mm.} This is in addition to choosing two values of magnetic shear. 
In the $y$-direction the grid is stretched so that the resolution is best at the neutral line $y=0$, with a minimum resolution of 0.0125 Mm. In the resultant simulations, the number of cells across the current sheet when it disrupts via non-linear tearing is about 80.

%The $y$ domain has a non-constant resolution, where the grid size is a cubic function of the $y$ index. The minimum cell size in $y$ is 0.0125 Mm and the maximum is 0.34 Mm. The initial current sheet of half width $a=0.5 ~ \textrm{Mm}$ is well resolved as there are 75 cells across the full width in the $y$ direction.

%Table \ref{tab:sims} shows the small parameter study performed. We vary both the length of the current sheet and the initial shear in the 3D magnetic field. As will be shown this leads to simulations with vastly different evolutionary paths.

As in \LDK, our numerical scheme will have a numerical resistivity much larger than the Sun's actual resistivity. Because we wish to study physical effects that depend on the resistivity, we must use an explicit value that is larger than the numerical value. We adopt $\eta=1.24\times10^{2} ~ \Omega.\textrm{m}$, or a diffusivity of $D=\eta/\mu_{0}=10^{8}  ~ \textrm{m}^2 \textrm{s}^{-1}$, as in the thought experiments above, and have verified that the current sheet diffuses  at a rate consistent with this value. %For a current sheet  half width ($a$) of $5\times10^{5} ~ \textrm{m}$ the diffusion time is $t_{d}=a^{2}/D = 2500 ~ \textrm{s}$, which, as will be shown, is longer than the tearing growth time, but still of the order of the duration of our simulation. Previous studies have mitigated this diffusion by removing the background diffusion of the original current sheet by inserting an additional electric field into the magnetic field evolution equation \citep[e.g.,][]{Del_Zanna_2016}. We do not perform such a  modification,  and so consistently follow the evolution of the very slowly diffusing current sheet as it becomes unstable.
For the chosen value of $D= \eta/\mu_{0}=10^{8} ~ \textrm{m}^2 \textrm{s}^{-1}$, and for the parameter realization of $L=100 ~ \textrm{Mm}$ and  $B_{x,0}=7.0 ~ \textrm{G}$, and defining the Alfv\'{e}n velocity in terms of the shear field, $V_{a} = B_{x,0}/\sqrt{\rho_0\mu_{0}}$, a typical Lundquist number $S=L V_{a}/D$ is $S=1.52\times10^{6}$, though the specific values at $t=0$ are given in the bottom two rows of Table \ref{table:calc}.

{\color{black} To force the current sheet to thin in the $y$ direction as a function of time, a slow (much smaller then both sound and Alfv\'{e}n speed) inflow
is induced by increasing gas pressure near the $y$-boundaries.  To avoid issues with large propagation speeds, first the magnetic field in the guide field direction in the region of driving is reduced using a smoothly varying function}
\begin{eqnarray}
    B_z(y) = B_z(y)_{orig}\left[1-\frac{0.75(1+\tanh{\frac{(|y|-y_{1})}{w_{1}}}}{2}\right]
\end{eqnarray}
{\color{black} such that for $|y|<y_1 = 5L_y/8$ the field is unchanged, and for $|y|>y_1$ the guide field is quarter of its original value ($w_1=L_y/32)$. We also increase the gas density in this region to ensure total magnetic pressure is maintained and the whole domain is still in equilibrium.
Then, to drive the flows by increasing the gas pressure, we increase the gas density using a forcing function in the continuity equation of the form (note that all variables are normalized)}
\begin{eqnarray}
    \frac{\partial\rho(t)}{\partial t} & = & \frac{\rho(t=0)}{\tau_{force}}\frac{t}{\tau_{ramp}}F(y), ~ t<\tau_{stop} \\
    & = & \frac{\rho(t=0)}{\tau_{force}}\frac{\tau_{stop}}{\tau_{ramp}}F(y), ~ t>=\tau_{stop}
    \label{eqn:force}
\end{eqnarray}
{\color{black} where the forcing term increases from $t=0$ to some time $\tau_{stop}=2300$, and is constant after that point. We set the gradient of this increase via $\tau_{ramp}=500$. The spatial function}
\begin{equation}
    F(y) = \frac{1+\tanh{(\frac{|y|-y_2}{w_2}})}{2}
\end{equation}
{\color{black} goes from zero away from the boundaries and around $|y|=y_2=6L_y/8$ increases to unity towards the boundary ($w_2=L_y/64$), so that the forcing is localized in the region near the boundary. The result of this dynamical forcing is the thinning of the initially thick current sheet, as discussed below and shown in the left panel of Figure \ref{fig:thinning_example}. For the 3D simulations discussed later, the forcing time $\tau_{force}$ is set to 10000, which results in a current sheet that thins with a timescale of approximately 1000 s (see Figure \ref{fig:thinning_example}). In the 2D parameter study in \S\ref{sec:2D}, this parameter is varied to produce different current sheet thinning timescales. }

\section{3D MHD Results}
\label{sec:3D}

\subsection{Parameter Study of Magnetic Shear and Current Sheet Length}

\begin{figure}  
\begin{center}
    \includegraphics[width=0.4\textwidth]{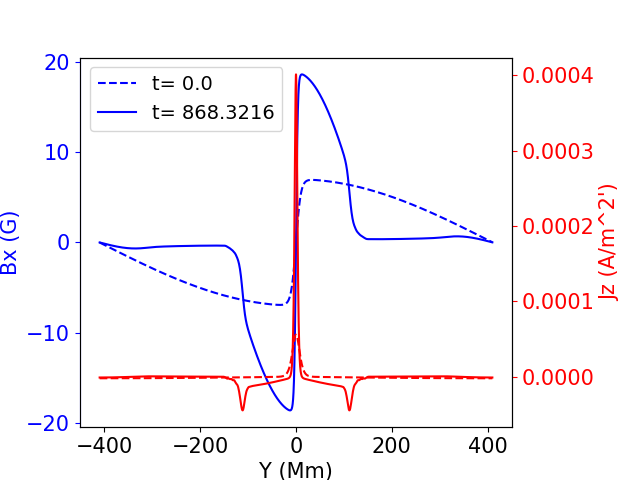}
    \includegraphics[width=0.4\textwidth]{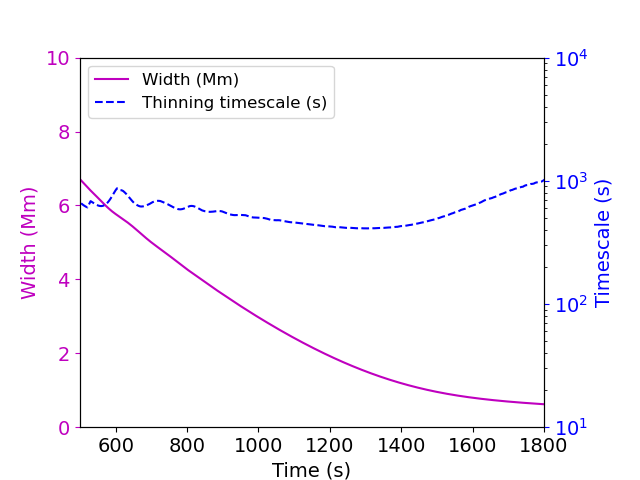}
    \caption{The effect of forced thinning: Left panel shows the shear field profile at two different times, and the $z$ component of the electric current density. The right panel shows the width of the sheet, obtained by performing a fit of the original profile to the instantaneous profile with variable $a$, $B_{x,0}$, and $B_{z,0}$. It also shows the thinning timescale, $\frac{a}{\partial a/\partial t}$. 
    \label{fig:thinning_example}}
    \end{center}
\end{figure}

We now show the results of four 3D MHD simulations. As mentioned above, we choose two values of $L_x=[20,100]$ Mm, and two values of $B_{x,0}=1.9, 7.0$ G. The values of $L_x$ are chosen from the theoretical studies above, called \ShortSheet and \LongSheet, respectively, based on the length of the dominant mode when the system starts to tear fast relative to the dynamical timescale, i.e. during the transition from $\tau_g>\tau_t$ to  $\tau_g<\tau_t$. The simulations evolve in much the same way as in \LDK, but now the current sheet is being constantly thinned with an approximate thinning timescale of 1000 s, see Figure \ref{fig:thinning_example}. 

The goal of these four simulations is to a) determine when, during the thinning, the 3D non-linearly evolving current sheet breaks up, and b) what are the dominant modes in the system when this happens? Recall that the answer to the latter question will determine if the system is in the \LongSheet regime and so is insensitive to the shear of the current sheet, or in the \ShortSheet regime and thus very sensitive to shear, as discussed in the Introduction.

\begin{figure}
\begin{center}
\includegraphics[width=0.4\textwidth]{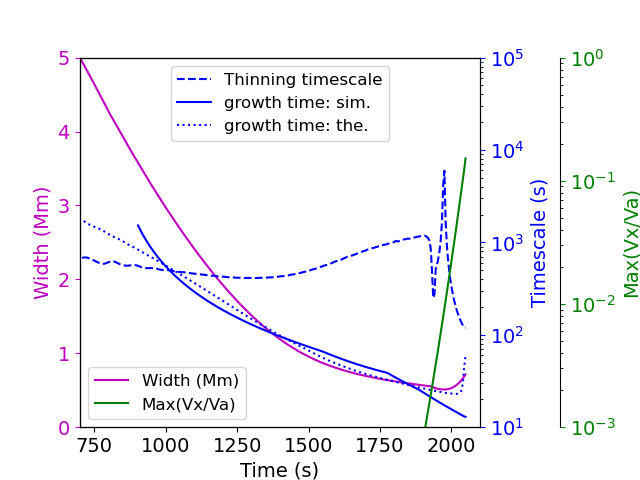}
\includegraphics[width=0.4\textwidth]{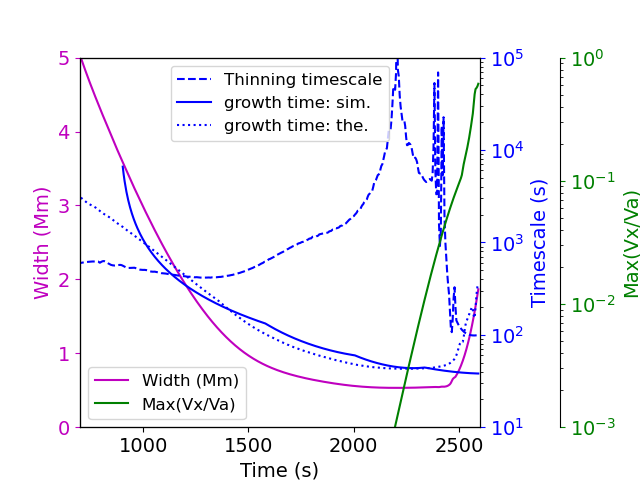} 
\\
\includegraphics[width=0.4\textwidth]{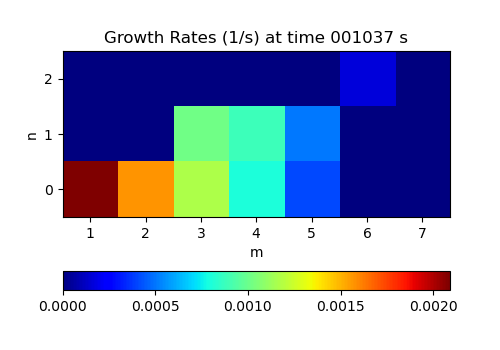}
\includegraphics[width=0.4\textwidth]{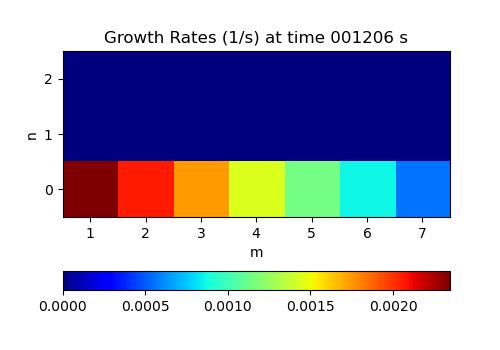}
\\
\includegraphics[width=0.4\textwidth]{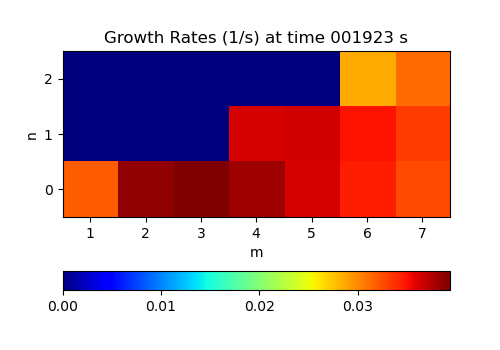}
\includegraphics[width=0.4\textwidth]{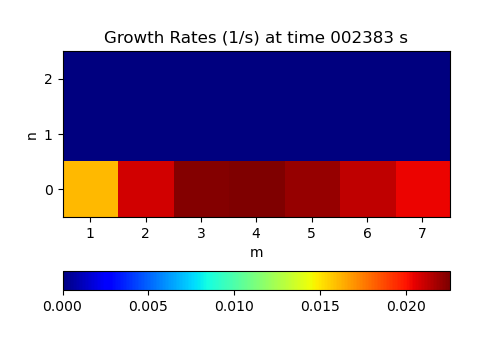}
\caption{Thinning 3D simulations with a long sheet (left panels: strong shear, right panels: weak shear). Top panels show the width, theoretical and simulated growth time of fastest tearing mode, thinning time, and maximum outflow velocity normalized by Alfv\'{e}n speed. 
Middle plots show the instantaneous growth rate of each mode when the fastest growth time equals the thinning rate. Bottom panels show when it is 1/50 of the thinning time. \label{fig:timescales_LS}}
\end{center}
\end{figure}

\begin{figure}
\begin{center}
\includegraphics[width=0.4\textwidth]{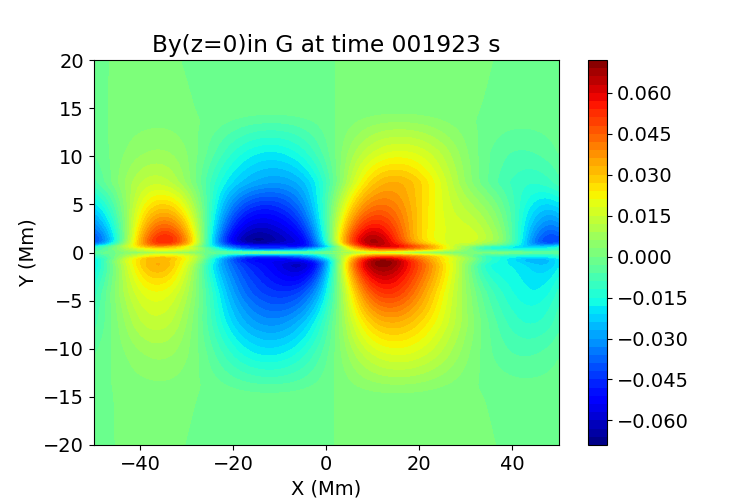} 
\includegraphics[width=0.4\textwidth]{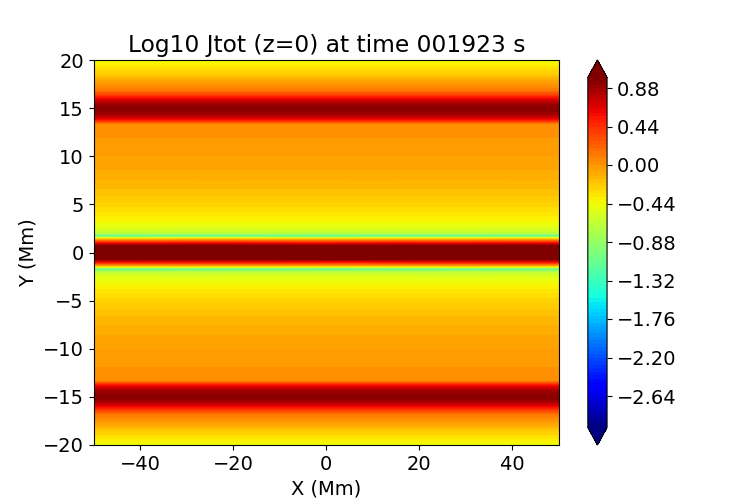} \\
\includegraphics[width=0.4\textwidth]{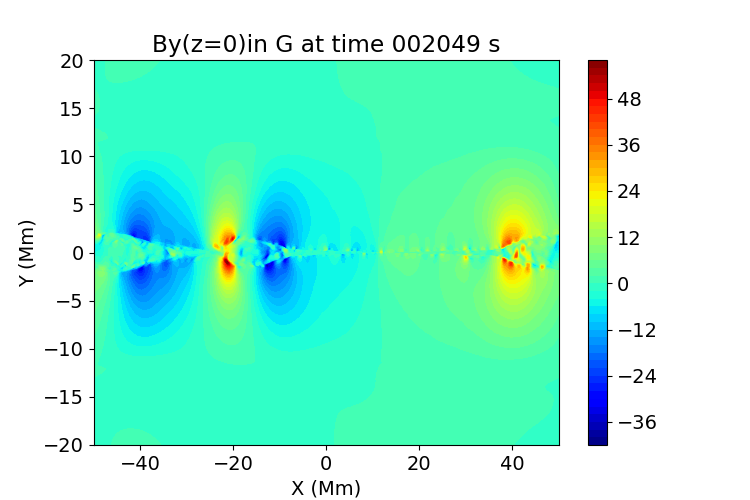} 
\includegraphics[width=0.4\textwidth]{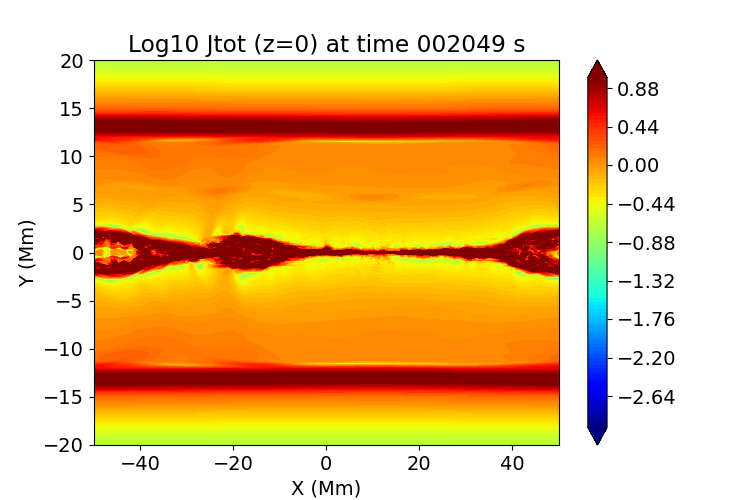}
\caption{Evolution of central current sheet for the strong shear long sheet case. The first time (top row) is when the growth time of the fastest tearing mode is 1/50 of the thinning time. The second time (bottom row) is chosen to show the late stages of coalescence of the $m>1$ parallel modes. \label{fig:images_2D_hs_LS}}
\end{center}
\end{figure}

\begin{figure}
\begin{center}
\includegraphics[width=0.4\textwidth]{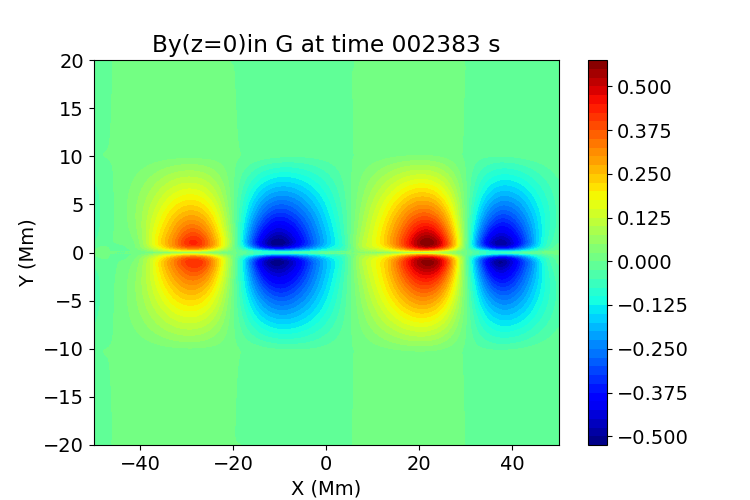}
\includegraphics[width=0.4\textwidth]{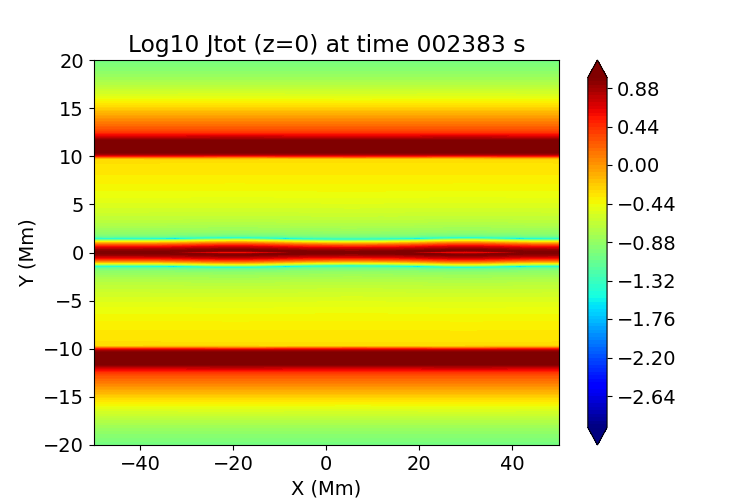} \\
\includegraphics[width=0.4\textwidth]{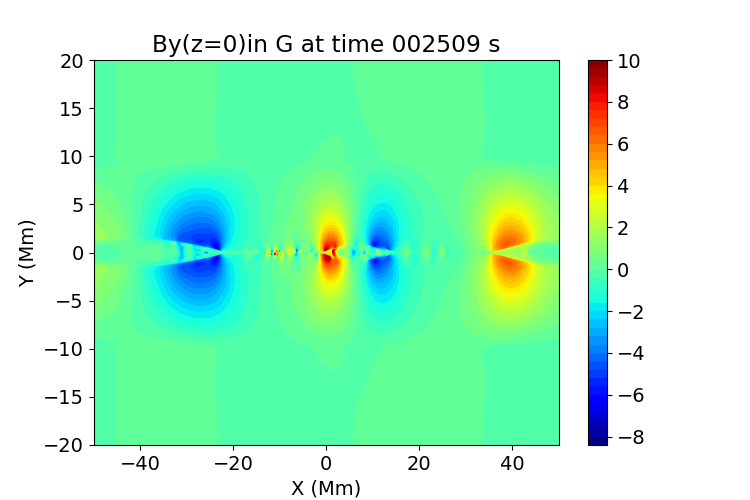}
\includegraphics[width=0.4\textwidth]{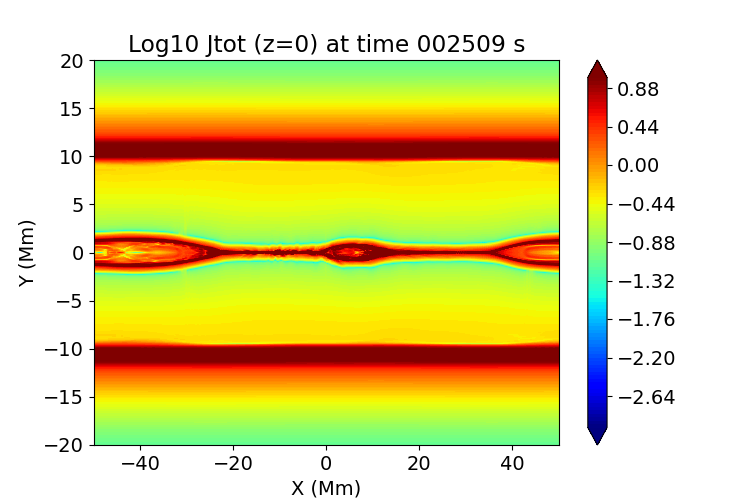}
\caption{Evolution of central current sheet for the weak shear shear long sheet case. The first time (top row) is when the growth time of the fastest tearing mode is 1/50 of the thinning time. The second time (bottom panels) is chosen to show the late stages of coalescence of the $m>1$ parallel modes.  \label{fig:images_2D_ws_LS}}
\end{center}
\end{figure}

{\color{black} As a result of the thinning, the magnetic field profile thins, and so the instantaneous values of current sheet width, density, and magnetic field in each direction change, all of which determine the growth rate of each mode based on linear theory. }{\color{black} To compare the growth of the instability in the simulation to the linear theory we do the following: At each time we fit the original 1D current sheet profile to the simulation data along the line $x=z=0$, which gives us an instantaneous value of $a(t)$, $B_{x,0}(t)$, and $B_{z,0}(t)$. We calculate the thinning timescale using $\frac{a(t)}{\partial a(t)/\partial t}$. These parameters are then used to calculate the theoretical growth rate of various modes using Eqn. (\ref{eqn:full_growth}). Then, from the simulation data, for each moment in time, we take Fourier transforms of the outflow, and for each Fourier mode we take its maximum value over the domain, and fit, for each mode, an exponential curve for a small window $[t-dt,t+dt]$ around the current time. We can then compare the simulated growth rate for each available mode to the theoretical value. Note that these values will be different from the idealized thought experiments in the previous section, as they idealized calculations assume an exponential thinning rate, rather than use the result of a 3D MHD simulation.}

{\color{black} This calculation is shown for the first two simulations with $L_x=100 $ Mm in the top panels of Figure 
%We first look at the two simulations with $L_x=100$ Mm. Figure 
\ref{fig:timescales_LS}. %shows, in the top panels. 
These panels show the thinning timescale, tearing timescale (inverse of the maximal growth rate) based on linear theory and simulation data, the width of the current sheet, and the maximum value of the horizontal flow normalized by the local Alfv\'{e}n speed. At this point  it is worth discussing how we characterize the system as ``non-linear". By the time that the current sheet can be seen to be visually deformed by the modes of the instability, it is likely already non-linear in that magnetic islands of each mode will have a width comparable to the inner reconnecting layer. We choose to use the outflow ($V_x$) created by the tearing modes, and define the start of the non-linear period when the maximum value of this component of the flow is 0.01 or 1\% of the local Alfv\'{e}n speed. As can be seen in the top two panels of Figure \ref{fig:timescales_LS} (strong shear simulation is on the left, and the weak shear simulation is on the right), as the width decreases, the tearing growth timescale goes from being slower than the thinning time to being faster. At some point later, the system goes non-linear, the outflows become a significant fraction of the Alfv\'{e}n speed, and the numerical fit of the current sheet width breaks down, leading to large oscillations in the plot of the thinning timescale. The first thing to note is that both simulations do go non-linear in terms of the outflows (green line). The breakup of the current sheet by the non-linear tearing instability thus happens in both strong and weak shear. The general evolution which shows this break-up can be seen in Figures \ref{fig:images_2D_hs_LS} (strong shear) and \ref{fig:images_2D_ws_LS} (weak shear), which show the reconnected field and the total current density on a log scale. }

%the maximum value of the outflow $V_x$ and fit this value to an exponential curve for a small window $[t-dt,t+dt]$ around the current time, this gives an estimate of the growth rate 
%At each time, we calculate the growth rate of each available parallel (n=0) and oblique (|n|>0) mode, based on this linear theory, as in \LDK. This requires a fit of the original profile with the values of $a(t)$, $B_{x,0}(t)$, and $B_{z,0}(t)$ as free parameters. We also calculate the growth rate in the simulations based on the value of the horizontal flow, which is driven by outflows associated with the various tearing modes, as in \LDK, wherein we showed a good agreement between linear theory and simulation. For the most dominant mode of the system at a given time, we then calculate the tearing timescale as the inverse of this growth rate.  The thinning time is calculated as $a(t)/(\partial a(t)/\partial t)$.}

We can understand the nature of the non-linear break-up of these two current sheets by looking at the dominant 3D tearing modes at different times, which are shown in the middle and bottom panels of Figure \ref{fig:timescales_LS}. We choose the cross-over time when $\tau_g=\tau_t$ and a later time before the non-linear break-up, when $\tau_g=\tau_t/50$.  The plots show the instantaneous theoretical growth rate of available modes at these times, for the strong shear (left) and weak shear (right).  Recall that, by design, for these two simulations with $L_x=100$ Mm, and based on the theoretical thinning study above, when the timescale of the instability becomes comparable to the thinning timescale, we expect the dominant mode's wavelength to be comparable to the box size (m=1), but that as the growth time decreases during the continued thinning, then higher wave-number (lower wavelength) modes become dominant, and so the system transitions to the \LongSheet regime where $m>1$. 

The middle panel of Figure \ref{fig:timescales_LS} shows that for both strong and weak shear, when $\tau_g=\tau_t$, the dominant linear mode is the (1,0) as in the theoretical thinning experiment. However, as we continue to when $\tau_g=\tau_t/50$, the dominant mode in the strong shear case is the (3,0) and in the weak shear case is the (4,0) mode. So indeed, both systems transition to the \LongSheet regime. Note that as in the static long sheet simulations of \citet{Leake2020}, the difference in strong and weak shear is the lack of additional oblique modes in the weak shear case.

Figure \ref{fig:images_2D_hs_LS} and \ref{fig:images_2D_ws_LS} show the evolution of $B_y$ and $J$ in the mid (z=0) plane in the strong and weak shear \LongSheet cases.  The snapshots are taken at the point in each simulation when $\tau_g=\tau_t$/50 and at a stage deemed to be non-linear, in that the maximum outflow velocity is 1\% of the local Alfven speed. In both simulations, the $\tau_g=\tau_t/50$ time shows a similar state. In both cases, the instantaneous dominant mode is either (3,0) or (4,0), but the $B_y$ patterns show that these modes have already started to coalescence and produce two large flux ropes. At the later time in each simulation, one can see the coalescence of these parallel modes and the break up of the current sheet. In both simulations,  {\color{black} the energy transfer from magnetic to thermal and kinetic} is driven by non-linear coalescence of the parallel modes, as in the static simulations of \citet{Leake2020}. 
This evolution shows that the amount of magnetic shear is not important for the onset of the reconnection in this long-sheet regime, as predicted, and seen in static simulations of \LDK. There is a slight delay in the weak shear case because the growth rates of the important modes are slightly slower. This is simply because of the ${B_x}^{0.5}$ dependence in Eqn. (\ref{eqn:tau_g}).

\begin{figure}
\begin{center}
\includegraphics[width=0.4\textwidth]{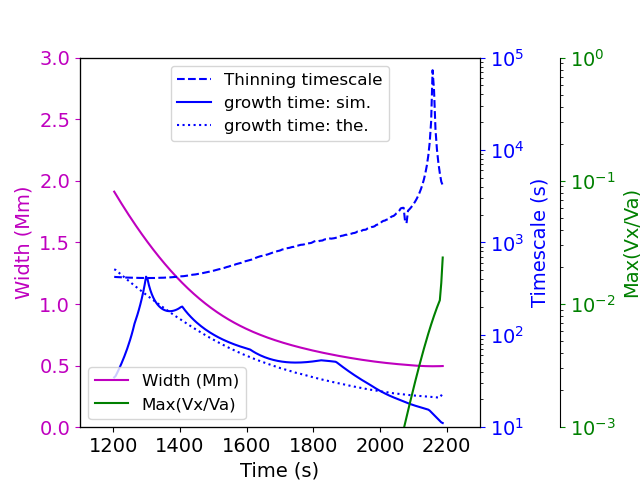}
\includegraphics[width=0.4\textwidth]{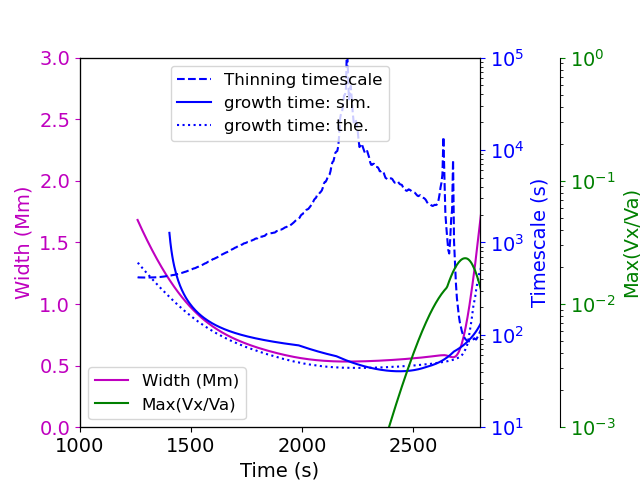}
\includegraphics[width=0.4\textwidth]{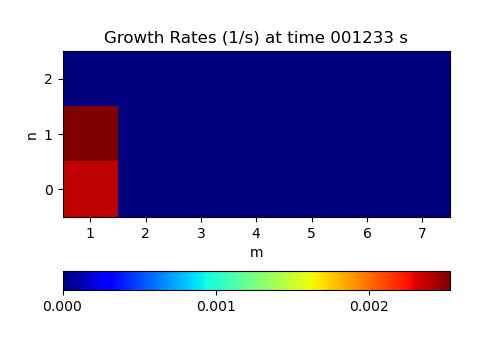}
\includegraphics[width=0.4\textwidth]{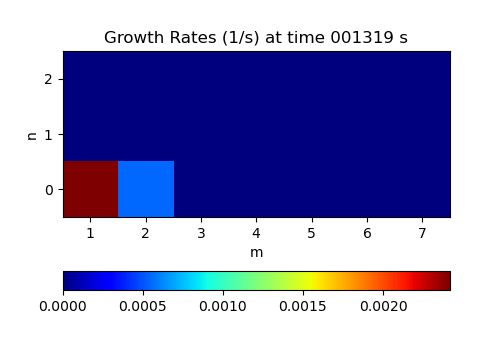}\\
\includegraphics[width=0.4\textwidth]{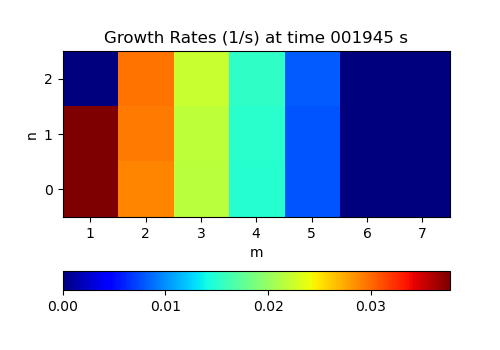}
\includegraphics[width=0.4\textwidth]{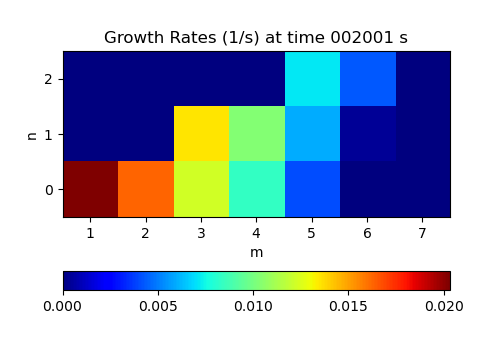}
\caption{Thinning 3D simulations with a short sheet (left panels: strong shear, right panels: weak shear)  Top panels show the width, theoretical and simulated growth time of fastest tearing mode, thinning time, and maximum outflow velocity normalized by Alfv\'{e}n speed.
Middle plots show the instantaneous growth rate of each mode when the fastest growth time equals the thinning time. Bottom panels show when it is 1/50 of the thinning time. 
\label{fig:timescales_SS}}
\end{center}
\end{figure}

\begin{figure}
\begin{center}
\includegraphics[width=0.4\textwidth]{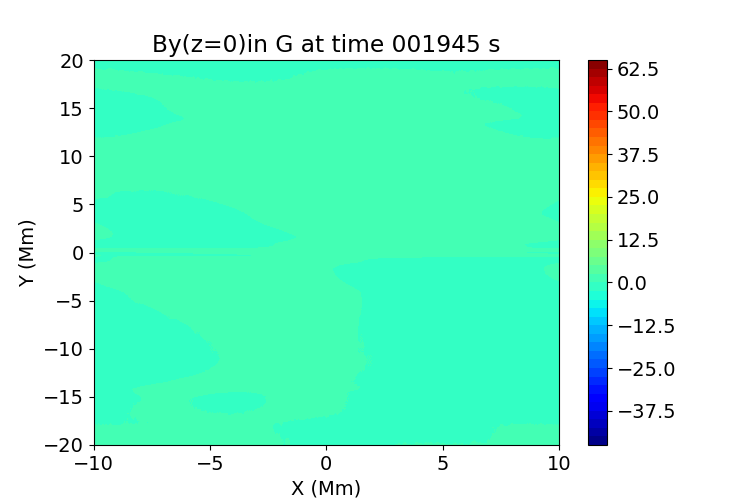} 
\includegraphics[width=0.4\textwidth]{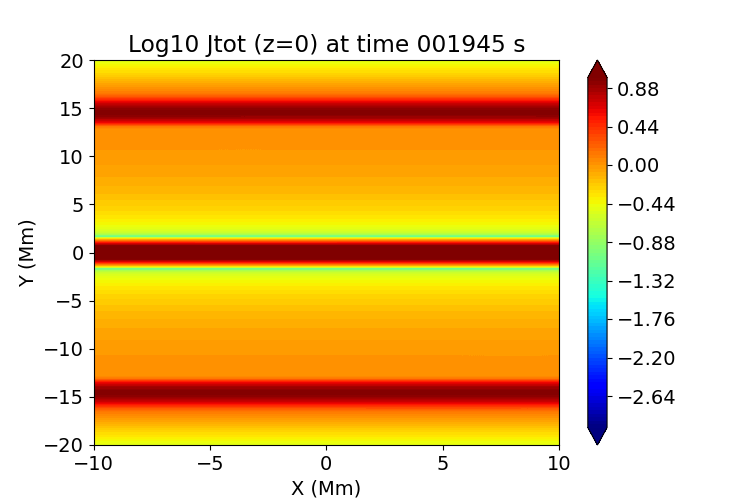} \\
\includegraphics[width=0.4\textwidth]{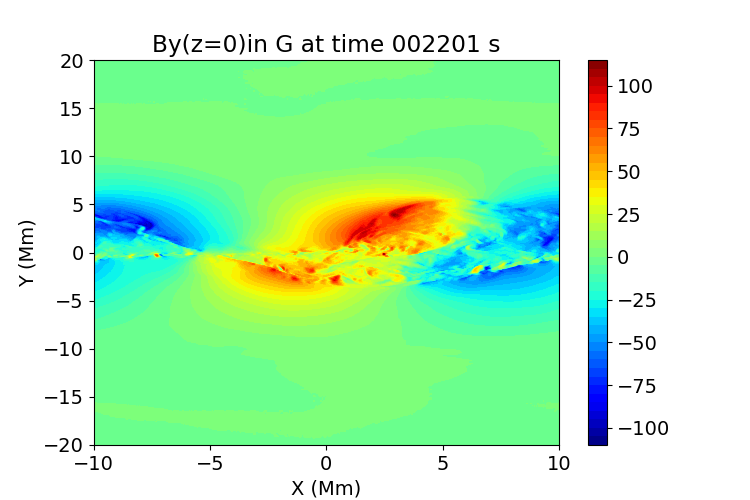} 
\includegraphics[width=0.4\textwidth]{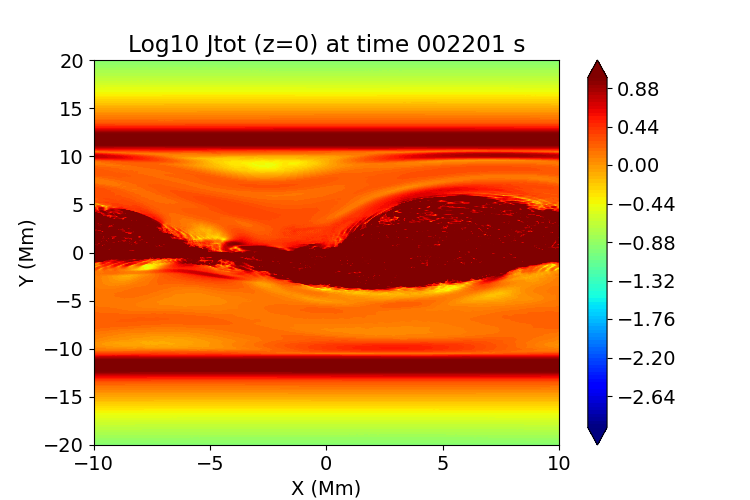}
\caption{Evolution of central current sheet for the strong shear short  sheet case. The first time (top row) is when the growth time of the fastest tearing mode is 1/50 of the thinning time. The second time (bottom row) is chosen to show the late stages of coalescence of the interaction of the (1,0) and (1,1) modes.    \label{fig:images_2D_hs_SS}}
\end{center}
\end{figure}

\begin{figure}
\begin{center}
\includegraphics[width=0.4\textwidth]{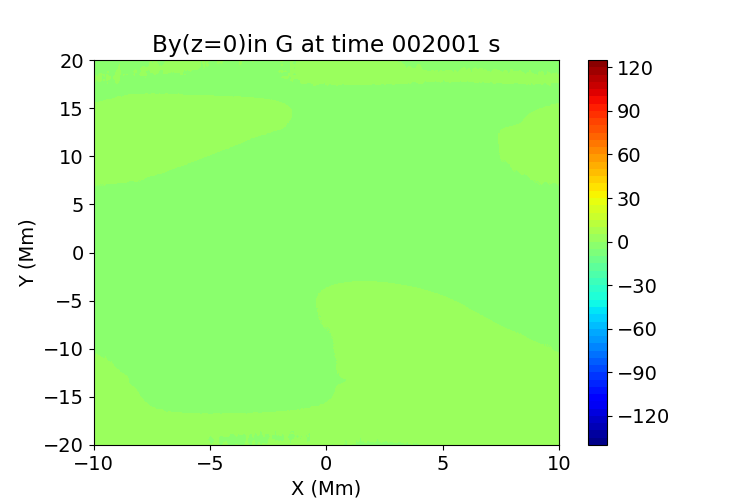} 
\includegraphics[width=0.4\textwidth]{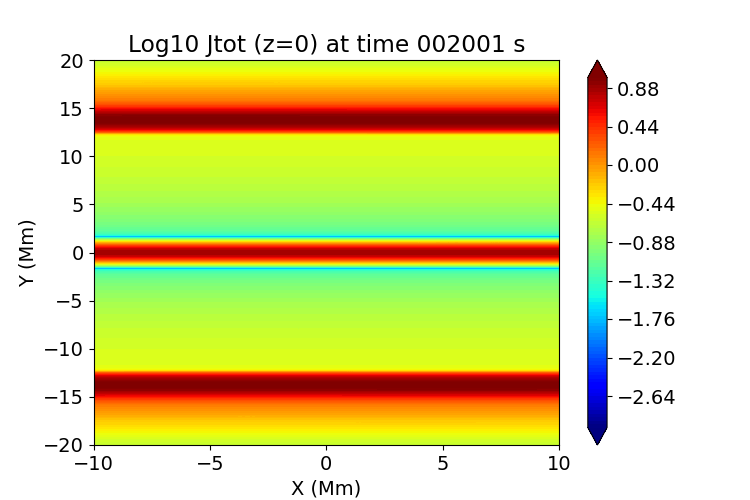} \\
\includegraphics[width=0.4\textwidth]{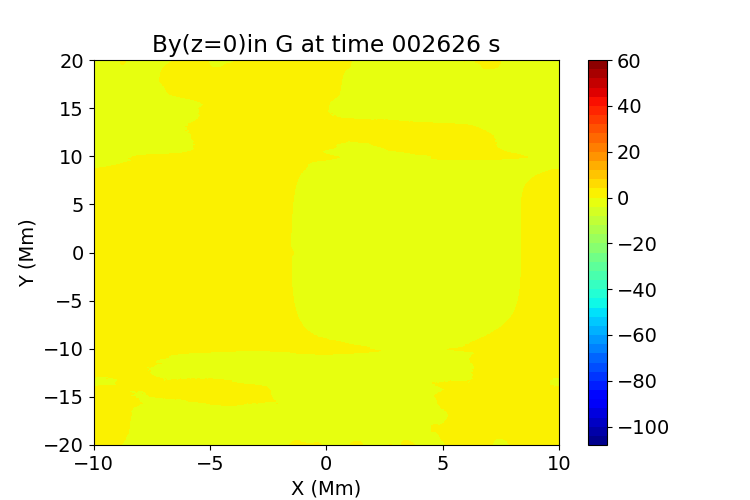} 
\includegraphics[width=0.4\textwidth]{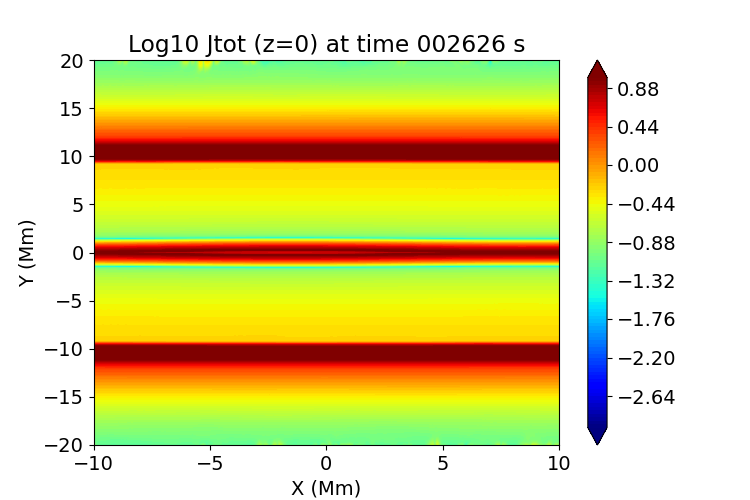}
\caption{Evolution of central current sheet for the weak shear short  sheet case. The first time (top row) is when the growth time of the fastest tearing mode is 1/50 of the thinning time. The second time (bottom row) is chosen to show the late stages of the saturation of the (1,0) mode.        \label{fig:images_2D_ws_SS}}
\end{center}
\end{figure}

Now we turn to the simulations with $L_x=20$ Mm. In the theoretical thinning experiment above, these thinning current sheets will remain in the ``short-sheet" regime until the point where $\tau_g << \tau_t$ and so should show similar behavior as the static short sheets in \citet{Leake2020}. 

Figure \ref{fig:timescales_SS} shows the analysis for the strong (left panels) and weak (right panels) simulations. In both cases, as the current sheet thins, $\tau_g$ decreases and transitions from being greater than to less than $\tau_t$, as shown in the top panels of Figure \ref{fig:timescales_SS}. As seen in the middle panels, the dominant parallel (n=0) mode at the point of cross-over is the (1,0) mode in both the strong and weak shear cases as predicted. However, one can see that in the strong shear case, there is a stronger oblique mode (1,1) which is not present in the weak shear case. 

 At a later time in the simulations with $L_x=20$, bottom panels of Figure \ref{fig:timescales_SS}, the dominant mode is still (1,0) in the weak shear case, and the (1,0) and (1,1) modes are strongest in the strong shear case.  So, in both  cases, the system is in the \ShortSheet regime, but in the strong case, the (1,0) parallel mode can interact with the (1,1) oblique mode. This is exactly the case as happened in the static simulations of \citet{Leake2020}. As a result of this non-linear interaction in the strong shear case, there is strong reconnection and the ratio of outflows to Alfv\'{e}n speed increases (see top left panel green line). However, in the weak shear case, the outflow (top right panel green line) seems to saturate and then decrease. In this thinning simulation with weak shear and a short-sheet situation, just as in the static cases of \cite{Leake2020}, the (1,0) mode saturates and the sheet does not break up.

This disparity in behavior when comparing the strong shear and weak shear in the ``short-sheet" regime is borne out when looking at the reconnected field ($B_y$) and current density ($J$) at an early and late time in both simulations. Figure \ref{fig:images_2D_hs_SS} shows these quantities for the strong shear case. At an early time, there are no strong signatures of any mode visible, although the (1,0) is dominant. At a later time, the (1,0) interaction with the (1,1) mode can be seen in both the $B_y$ and $J$ plots. Figure \ref{fig:images_2D_ws_SS} shows the same for the weak shear case, and the (1,0) saturation can be seen at later times, with no disruption of the current sheet. Note the enormous difference in the magnitude of $B_y$.

These four 3D simulations have borne out the predictions made based on the theoretical thinning experiment in \S\ref{sec:theory}. Firstly, the current sheet only breaks up once $\tau_g$ becomes less than $\tau_t$, i.e. when the growth of the tearing instability is happening faster than the thinning or dynamical timescale of the system. Secondly, the behavior of the system in terms of the dependence on shear is the same as the static simulations of \LDK, but now is playing out in a dynamically thinning situation. For \LongSheet simulations, the shear has only a weak effect on the evolution, and the sheet breaks up as higher m  parallel modes coalesce. For \ShortSheet simulations, only the strong shear case breaks up when the (1,0) and (1,1) modes interact non-linearly.  The weak shear saturates with the (1.0) mode unable to disrupt the current sheet. We propose, that as the \ShortSheet simulations here are applicable to the situations in nanoflares, this dependence on shear in the \ShortSheet regime is a possible explanation for the ``Parker angle" important for understanding coronal heating \citep{Klimchuk_2015}.

\section{Two-Dimensional Parameter Study of Resistivity and Thinning Rate}
\label{sec:2D}

The above 3D results apply to a chosen diffusivity and thinning rate. The former of these two parameters is significantly larger than that of the Sun, as discussed in \S\ref{sec:theory} and \S\ref{sec:numerics}, although the simulations here do lie in the same regime as the Sun in terms of the ratio of dominant mode wavelength to current sheet length. Ideally, one would perform a parameter study of the diffusivity. However, these 3D simulations are too computationally intensive for such an approach.

Luckily, the fact that the long sheet simulations behave in a similar way for both strong shear and weak shear means one can use 2.5D simulations to perform a parameter study. This is because the reduction of the problem to 2.5D is effectively the limit of reducing the magnetic  shear, or effectively, increasing the guide field. In 2.5D, there are no oblique modes, just as in the very weak shear case. So in these 2D simulations, we use the strong shear value of $B_x=7$ G.

The questions that this parameter study will address are: a) Does the onset always occur after the cross-over point, when $\tau_{g}<\tau_{t}$ when we use different thinning rates and diffusivities? and  b) Is there a predictable delay from this cross-over to the onset of the break-up of the current sheet by the non-linear tearing instability? This latter question has practical applications: If one can know the diffusivity, Alfv\'{e}n speed,  thinning rate, and dimensions for a given observed current sheet, one could predict when the current sheet will breakup, and hence how much free energy can build up and is available to be released (see discussion in \S\ref{sec:discussion}). This has important consequences for the thermalization of energy and the heating of the closed corona.

This parameter study uses three values of diffusivity, of $10^8$, $10^{8.5}$, and $10^9$ $\textrm{m}^2/\textrm{s}$. The lower value is dictated by the numerical resistivity of the code, which is dependent on the resolution used. Of course, we could go to a lower resistivity, but this would require better resolution and this would reduce our ability to perform a parameter study. The upper value is dictated by the diffusion of the current sheet at higher resistivities which counters the thinning process.

We also use three different thinning rates. {\color{black} As discussed in \S\ref{sec:numerics}, we add a source term in the continuity equation  near the $y$ boundaries to drive an inflow, and we can vary the parameter $\tau_{force}$ in Eqn. (\ref{eqn:force}) to produce different thinning rates. We choose values of $\tau_{force}=1000,10000,100000$ to give approximate thinning timescales of $(t_1,t_2,t_3)=(1000,500,300)$ s. Note that the actual thinning rate will vary in time around these values as the simulations develop,  as was shown in Figures \ref{fig:timescales_LS} and \ref{fig:timescales_SS} for the $t_2$ 3D case. }

{\color{black} As discussed in \citet{Comisso2017}, the initial noise can also affect the evolution and onset of fast reconnection in tearing sheets. However, our initial perturbation is very small (average m/s) and so we propose that the dominant effect on the time evolution is the dynamical thinning. We performed some initial tests with weaker perturbations at $t=0$, and found little diference in the delay time calculations here. }

\begin{figure}
    \begin{center}
        \includegraphics[width=0.45\textwidth]{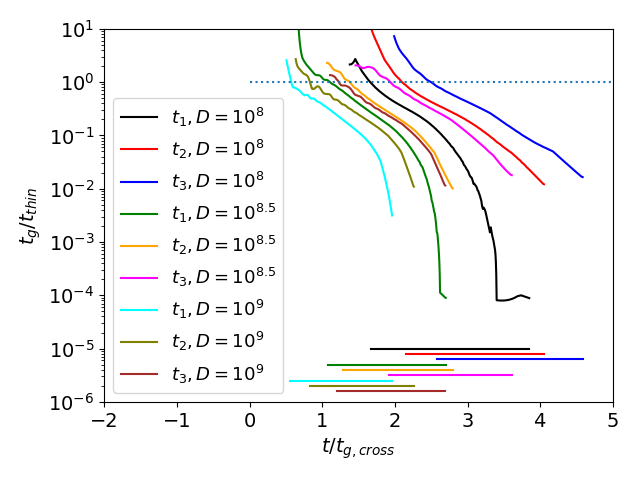}
        \includegraphics[width=0.45\textwidth]{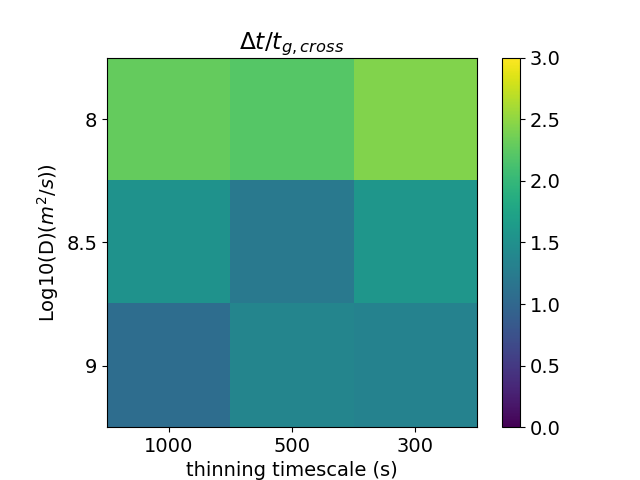}
    \end{center}
    \caption{Left panel: Ratio of the growth timescale of the fastest mode, $\tau_g$, to the thinning timescale $\tau_t$, as a function of time, where the time is normalized by the growth time when $\tau_g=\tau_t$, for all 9 simulations in the 2D parameter study. The horizontal lines indicate the delay on the $x-axis$ from when $\tau_g=\tau_t$ to the onset of the non-linear stage of the tearing instability and thus the break-up of the current sheet. The right panel shows this delay as a function of thinning time and diffusivity.
  \label{fig:2D_param}}
\end{figure}

Figure \ref{fig:2D_param} shows the general results of the parameter study. In each simulation, as before, we calculate the instantaneous growth rate for each mode using the Fourier transform of the outflows, and then work out the tearing growth timescale by taking the inverse of the growth rate of the fastest growing mode. To allow all the simulations to be plotted on the same panel, we then divide this by the instantaneous thinning timescale in each simulation. This is shown in the left panel of Figure \ref{fig:2D_param}, as a function of time. The time is also normalized by the value of $\tau_g$ when $\tau_g=\tau_t$. The plots are stopped when the outflows become 1\% of the local Alfv\'{e}n speed, as this was prescribed to be the non-linear stage of the tearing instability. It is worth noting that all 9 simulations transition from $\tau_g>\tau_t$ to $\tau_g<\tau_t$ BEFORE the system is disrupted, as predicted.

The delay from the point when $\tau_g=\tau_t$ to the non-linear point is calculated for each of the 9 simulations. This delay is shown for all 9 simulations on the left panel of Figure \ref{fig:2D_param}, as horizontal lines. This delay is shown as a function of the thinning timescale and diffusivity in the right panel. There is no obvious relationship between the delay and the parameters but it is worth noting that the range in delay, normalized by $\tau_g(\tau_g=\tau_t)$ is small (between 1 and 3 multiples of $\tau_g$).

This small 2D parameter study shows that indeed, the onset of the breakup of the current sheet occurs AFTER the point when the thinning timescale decreases from ABOVE the thinning timescale to BELOW the thinning timescale.  Furthermore, the breakup of the current sheet occurs 1-3 growth times after this point. This allows an approximate prediction of the breakup of a dynamically thinning current sheet. In particular, knowing the Alfven speed, diffusivity, and thinning timescale, we can estimate the width when the thinning and growth times are equal and also $\sim 2$ growth times later, when breakup occurs.

\section{Discussion}
\label{sec:discussion}
We have performed multi-dimensional simulations of the onset of the tearing instability in a dynamically evolving current sheet. 

Initially we performed a thought experiment of thinning current sheets relevant to coronal conditions. We imagined a thinning current sheet with a known thinning rate, initial width, density,  and magnetic field profile. We were then able to estimate the growth timescale of the tearing instability as a function of time as the system thins, and calculate when the growth timescale of the instability crosses from being slower than to faster than the thinning timescale. We were also able to estimate when the growth timescale became a small (1/50) fraction of the thinning timescale. At these points in time we were also able to see what the dominant 3D tearing modes of the system were, and place the system into one of two categories. In one case, the fastest growing mode's wavelength was less than the length of the current sheet, or ``long-sheet" regime. In the other case, the fastest growing mode's wavelength was more than the length of the sheet, or ``short-sheet" regime. We confirmed that the eruptive flare situation in the solar corona lies in the long-sheet regime, and the nanoflare situation lies in the short-sheet regime. We also confirmed that our planned 3D parameter study, using lengths of $L_x=20$ Mm and $L_x=100$ Mm would allow us to investigate both regimes, despite our diffusivities being very different from the values in the solar corona.

We then performed four 3D MHD simulations of a dynamically thinning current sheet. The thinning was driven by pressure gradients near the boundaries generated by a steady increase in density. This thinning transitioned the system from the growth time of the instability being slower than the thinning time to it being faster. At some point after this transition, the current sheet broke up as the tearing instability became non-linear. By design, we had two simulations that lay in the "long-sheet" regime when the system became non-linear, one with strong shear and one with weak shear. We also had two simulations that lay in the "short-sheet" regime, with the same two values of shear. 

Firstly, we note that the current sheet only breaks up once $\tau_g$ becomes less than $\tau_t$, i.e. when the growth of the tearing instability is happening faster than the thinning or dynamical timescale of the system. {\color{black} This is consistent with the 2D simulations of \citet{Tenerani2015}.}

One major result from our previous work in \citet{Leake2020} was that in the long-sheet regime, the main effect of magnetic shear in the sheet was the strength of the oblique modes (strong shear means stronger oblique modes) but that the non-linear stage was dominated by coalescence of the higher wave-number parallel modes, which eventually merged to one large flux rope. Thus the magnetic shear was not important in the long-sheet regime for the onset of the reconnection. 
Conversely, in the short-sheet regime, the dominant parallel mode was the (1,0) mode, and so there was no coalescence between multiple parallel modes. However, for strong shear the (1,1) oblique mode was able to interact non-linearly with the (1,0) parallel mode and drive significant outflows. For weak shear the oblique modes are too weak to drive interaction with the (1,0) mode and so this parallel mode saturates with no strong outflow. Thus magnetic shear was a very important parameter in the short-sheet regime, potentially being a switch-on parameter that would determine when the energy is released, and related to the well-known Parker angle.

%This general pattern, of shear being unimportant in the \LongSheet regime, but very important in the \ShortSheet regime, plays out in the 3D dynamically thinning simulations in this paper. After the transition to $\tau_g<\tau_t$, whether the system was in the long or short-sheet regime, and whether the shear was strong or weak, determined how it would behave non-linearly. \LongSheet simulations broke up for both weak and strong shear, but only the \ShortSheet strong shear simulations showed a break-up of the current sheet. 

 We propose, that as the \ShortSheet simulations here are applicable to the situations in nanoflares, this dependence on shear in the \ShortSheet regime is a possible explanation for the "Parker angle" important for understanding coronal heating \citep{Klimchuk_2015}. Using only estimates for the linear evolution, given estimates of the current sheet structure (field, width, density), we can predict when it will break up and how it will break up. This has consequences for theoretical studies of the release of free magnetic energy during eruptive and non-eruptive events in the corona, and for observations of these eruptive events. 

Finally, to show that these results are not just specific to one choice of thinning rate and diffusivity we performed a small parameter study in 2D, varying the diffusivity and tearing rate by about one order of magnitude each, but using the long sheet strong shear case to ensure we observed a breakup of the current sheet. In all cases, the current sheet broke up due to significant reconnection only after the transition from $\tau_g>\tau_t$ to $\tau_g<\tau_t$ and the delay from this transition to non-linear breakup showed only a weak variance within the parameter space. This provides some confidence that the predictive approach described above can be applied to the solar corona. 

These results make it possible to estimate when a current sheet will break up and therefore how much magnetic energy can build up in a field and be available to power coronal heating, coronal mass ejections, flares, etc. Consider a uniform initial field where $B = B_{z0} =$ constant. Suppose the field is subjected to a simple shear flow of scale $a_0$. A current sheet forms with this thickness. The shear component of the field outside of sheet, $B_x(t)$ increases with time, as does the total field
\begin{equation}
B(t)^2= B_{z0}^2 + B_x(t)^2 .
\end{equation}
Only the guide field component is present at the center of the sheet, $B_{z,c}(t)$. It maintains its initial value, $B_{z0}$, at the line-tied boundaries where the driving is applied, but it decreases with time away from boundaries as the magnetic pressure increases and squeezes the sheet. Magnetic flux is conserved during the squeezing:
\begin{equation}
B_{z0} a_0 = B_{z,c}(t) a(t) .
\end{equation}
Pressure balance across the sheet requires that 
\begin{equation}
B_{z,c}(t)^2 = B(t)^2 .
\end{equation}
Combining, we obtain
\begin{equation}
B_x(t)^2 = B_{z0}^2 \left[ \left( \frac{a_0}{a(t)} \right)^2 - 1 \right].
\end{equation}
The free energy in the field, $B_x(t)^2/8\pi$, can thus be estimated if we know the field strength, the initial thickness of the sheet (the scale of the driving), and the current sheet thickness at the time of reconnection. Note that when $a(t)$ is small compared to $a_0$, i.e., after there has been substantial thinning, $B_x^2 \propto a^{-2}$, so the energy has a strong inverse dependence on sheet thickness.

These 3D MHD simulations are obviously simplified compared to the solar corona. Current sheets are not infinite in extent in $x$ and $z$ and so we must consider the effect of line-tying in these directions. This is addressed in an upcoming paper \citep{Daldorff2024}.

%Magnetic shear as a switch-on parameter for current sheets finite in length (not just width) is investigated in an %upcoming paper \citep{Klimchuk2024}. Future work will combine these approaches and look at dynamically evolving current %sheets which are finite in extent in all dimensions. 

In this paper we have discussed how the onset of reconnection occurs when thinning current sheets have reached the scale where the growth rate of the fastest tearing mode exceeds the thinning rate. This is not the only possibility. In another study we have concluded that evolving current sheets can reach critical conditions where equilibrium is no longer is possible \citep{Klimchuk2023}. They then spontaneously collapse and reconnect. The critical conditions are
\begin{equation}
\frac{B_x}{B_{z,0}} > \frac{2 \lambda}{L} ,
\end{equation}
where $B_x$ and $B_{z,0}$ are respectively the shear and guide field components outside the sheet, $\lambda$ is the sheet length, and {L} is the separation of the line-tied footpoints along the field. We are presently testing our theoretical prediction with numerical simulations and will report our results shortly. 

Shearing by driver flows - as occurs commonly on the Sun - leads to reconnection onset in both scenarios. Such driving obviously increases the ratio of shear to guide field components. It also increases the magnetic pressure, which causes the sheet to thin. Which onset mechanism occurs first  depends on the details of the situation, including the speed of driving. We will explore realistic possibilities in the near future.

\acknowledgements

JEL, LKD, and JAK wer funded by NASA's Internal Scientist Funding Model (ISFM), and NASA's Living with a Star (LWS) Program. Simulations were performed on NASA's High End Computing Facilities. Source code, raw data, and plotting routines are all available upon request from the authors.

\bibliography{bibliography1, bibliography2, bibliography3, bibliography4,bibliography5}

\end{document}